# An empirical study of Unfairness and Oscillation in ETSI DCC

Fatma Marzouk[1]    Rachid Zagrouba[2]    Anis Laouiti[3]    Paul Muhlethaler [4]    Leila Azzouz Saidane[5]

[1] RAMSIS Team, CRISTAL Laboratory,Manouba, Tunisia. Email: fatmarzouk@gmail.com
[2] Higher Institute of Computer Science, Ariana, Tunisia. Email: rachid.zagrouba@cristal.rnu.tn
[3] TELECOM SudParis, RS2M,France. Email: anis.laouiti@telecom-sudparis.eu
[4] INRIA, Paris-Rocquencourt, France. Email : paul.muhlethaler@inria.fr
[5] National School of Computer Science, Manouba, Tunisia. Email: leila.saidane@ensi.rnu.tn

*Abstract*—**Performance of Vehicular Adhoc Networks (VANETs) in high node density situation has long been a major field of studies. Particular attention has been paid to the frequent exchange of Cooperative Awareness Messages (CAMs) on which many road safety applications rely.**

**In the present paper, se focus on the European Telecommunications Standard Institute (ETSI) Decentralized Congestion Control (DCC) mechanism, particularly on the evaluation of its facility layers component when applied in the context of dense networks. For this purpose, a set of simulations has been conducted over several scenarios, considering rural highway and urban mobility in order to investigate unfairness and oscillation issues, and analyze the triggering factors.**

**The experimental results show that the latest technical specification of the ETSI DCC presents a significant enhancement in terms of fairness. In contrast, the stability criterion leaves room for improvement as channel load measurement presents (*i*) considerable fluctuations when only the facility layer control is applied and (*i.i*) severe state oscillation when different DCC control methods are combined.**

*Keywords—component; CAM, DCC, VANET, ETSI-ITS*

## I. INTRODUCTION

In recent years, the interest in using Intelligent Transportation Systems (ITS) has grown considerably. A wide range of vehicular ITS applications are based on the periodic exchange of Cooperative Awareness Messages (CAMs). In such messages, vehicles periodically send information related to their own position, speed and heading [1]. Receiving CAMs from its neighbors should provide a node with an accurate assessment relevant image of its neighborhood. However several studies [2][3] have pointed out that the channel can quickly become congested due to the massive exchange of such status messages. This situation may worsen as VANETs become increasingly common.

Many standardization activities have been carried out by ETSI in Europe and IEEE in North America in order to specify how the channel should be accessed and to establish rules regarding the frequency and dissemination of periodic safety messages.

In terms of physical (PHY) and medium access (MAC) specifications, both standards are based on IEEE802.11p [4]. While WAVE uses a fixed 10Hz frequency for disseminating CAMs during the slots dedicated to safety messages as specified by the alternating access scheme [5] with the enhanced distributed channel access (EDCA), the European standard follows a different approach for generating CAMs with a frequency ranging from 1Hz to 10Hz depending on the mobility conditions of the vehicles and the current channel load, as will be detailed in this paper.

In addition, IEEE WAVE and ETSI ITS G5, can support congestion control algorithms [6] and enhancements so as not to compromise the application requirement level.

Our work aims to analyze the behavior of a real-life dense network scenario where the DCC facility layer component has been implemented, in order to study unfairness and oscillation issues that have been reported in the latest technical specifications of the DCC mechanisms.

## II. RELATED WORK

The congestion control problem for safety message traffic was initially tackled using three different strategies. The first one consists in controlling the transmission power among nodes so that each vehicle can perceive a less occupied channel. One of the proposals put forward in [7][8] is to dynamically and fairly control the transmission power and to piggyback the transmission power in beacon messages in order to make the global channel load converge to a target threshold. While the effectiveness of the algorithm depends on the beacon frequency used, the overhead induced cannot be neglected.

The second class aims to adapt the transmission rate of CAMs messages. In [9] the authors present an adaptive approach for rate adjustment to reach a target based on a mathematical expression is presented, showing. They provide results in terms of adaptation to the network's dynamic and convergence to the targeted rate. In contrast, in [10] the rate is adapted according to the number of neighboring vehicles.

The last category contains algorithms and frameworks where hybrid adaptation methods are used. For instance, in [11] rate control is firstly performed, then it is followed can by power control once the minimal beacon transmit frequency has been reached and the channel load remains high.

As regards congestion, DCC is characterized by cross-layer architecture which is implemented within the ITS-Station. The core of DCC is a finite state machine composed of 3 states (Relaxed, Active, and Restrictive). In each state, the DCC components set different values for the station parameters: DCC-Access [12] acting on transmission Power, transmit Rate, and CCA sensitivity threshold. DCC-Facility operates on CAM [13] and DENM. DCC-Management [14] operates as a cross layer entity. Finally, the DCC-Network provides the vehicle with global information (DCC parameters Received from neighboring vehicles). That is important for parameters evaluation.

Several studies have focused on the evaluation of DCC. In [15] the authors conclude that it induces a considerable local and global oscillations in channel load measurements. Similarly, in [16] where the same observation was explained by the lack of simulation resources to reach a high CL sampling rate. In addition simulations in [17] show through a set of simulation a remarkable unfairness problem among neighboring node. The divergence between these two results is due to by the two different strategies of extracting the channel load (CL). While in the first one, accessing the CL measurement is done at a synchronized time for all nodes, the procedure is done asynchronously in the third study. The second approach is feasible as the nodes' measurements can be insured by tracking the GPS time. So it is obvious that the second approach is more rational as the vehicles will have different CAM transmission rate (time?). The Unfairness problem was tackled in [18] and [17] by piggybacking channel load measurements and the state of each node in each beacon sent. Furthermore, the reported oscillation problem occurred in scenarios where at least 3 DCC parameters were regulated.simultaneously

### III. DCC FACILITY LAYER

The specification of the DCC CAM facility component stipulates that a vehicle generates CAMs under to two main conditions: First, if the time elapsed since the last CAM generation is equal to or greater than a period set by the DCC state machine $T\_GenCam\_Dcc$ and one of the ITS-S dynamics related conditions is triggered i.e the absolute difference in heading, distance or speed included in previous CAMs exceeds 4°,4 m,and 0,5 m/s respectively. Due to the second condition, the time elapsed since the last CAM generation will be equal to or greater than the period representing the currently valid upper limit of the CAM generation interval $T\_GenCam$ and equal to or greater than $T\_GenCam\_Dcc$.

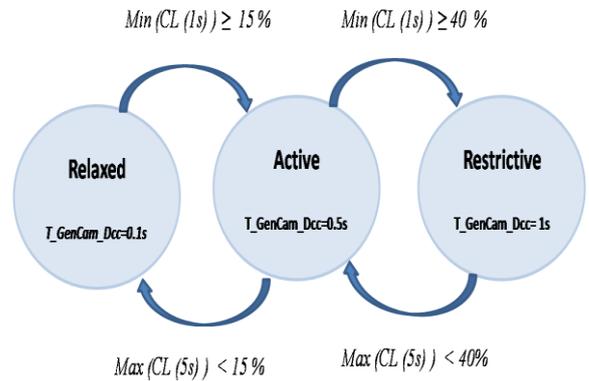

Fig.1 DCC state Machines

The CAM generation mobility condition is checked after $T\_GenCam\_Dcc$ at least in order to trigger a sending event if valid according to condition 1).otherwise, the transmission occurs according to condition 2) at the latest 1s after at the last send time. Fig 1 shows the $T\_GenCam\_Dcc$ depending on the current finite state of the vehicle. The latest version of the standard [13] advocates a value of 0.1 s as the shortest CAM generation period for the relaxed state rather than 0.04s in previous version. It also specifies that, after triggering a number $N\_GenCam$ of consecutive CAMs due to condition 2), $T\_GenCam$ will be set to 1s. As shown in Figure.1 transitions from a state $i$ to state $i+1$ can take place if the minimum channel load during 1 second exceeds a corresponding threshold. The reverse transitions may occur after a period of 5s.

### IV. EVALUATION EXPERIMENTS

The Ns3 simulator was used for the implementation and evaluation of the DCC facility layer. It is an open-source event based simulation environment written in C++, which provides a number of useful features. A rural highway scenario and an urban scenario with large number of intersections were selected. In order to provide a realistic case study, we used to SUMO [19] to set the vehicular traffic over a snippet of the two-lane highway linking the cities of Sousse and Sfax (Tunisia) and an urban area located in the town of Ariana (Tunisia). Figure 2 shows the map of the highway, downloaded from OpenStreetMap (OSM) and edited using Java OpenStreetMap Editor (JOSM). Figure 3 shows the urban region.

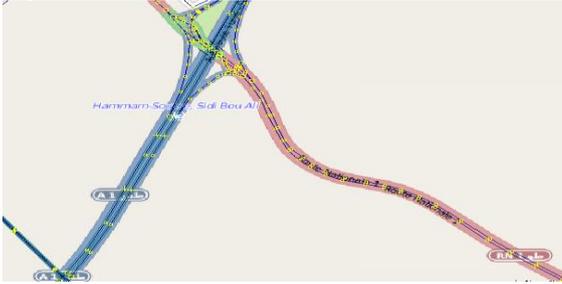

Fig.2 Rural highway scenario: Sousse-Sfax highway

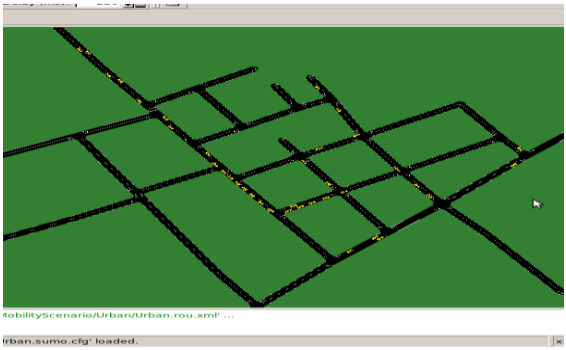

Fig.3 SUMO Simulation of the urban area in Ariana

TABLE I SIMULATION SETTING

| Parameter | Value |
|---|---|
| Tx power | 27 dBm |
| CCA threshold | -125 dBm (CCA Range 1.3km) |
| Tx Rx Gains | 1 dBm |
| CBR sampling rate | 1 khz |
| Propagation model | Log distance (Loss exponent m=3) |
| CAM size | 600 Bytes |
| Number of vehicles | 600 |
| Speed | [25-44km/s] |
| Jitters | 2s /4s/ 6s /12s |

In both scenarios the region size is equal to $1.3km \times 1.3km$. We studied fluctuations in a dense region with 600 vehicles in both scenarios. Because we are mainly interested in the impact of urban and rural mobility, we used a long-distance path loss model with an exponent of 3 for both scenarios to simulate simple radio propagation, without taking into account most of the PHY layer effects such as shadowing, multipath propagation, etc. The simulation parameters are summarized in Table I.

The first set of simulations aimed to investigate the existence of the reported DCC unfairness between nodes. As already discussed in the previous section, this situation might be encountered when the vehicles have different times to read CL measurements and that in the network was subject to oscillation. As perfect synchronization in CL measurement time wouldn't reflect the real channel occupation (sending CAM messages occurs at different times) this case is excluded from the scope of our study. Thus, to examine the two situations, we applied different jitter values to the nodes so that the time that their first CAM was sent belonged in the interval of [0, jitter value]. We started with relatively small values of jitter. Then we measured the effect of raising the value. Over 300 seconds of simulation using values of 2s, 4s, 6s, 12s ,25s and 50s no divergence between the nodes was observed. Performing another simulation using the same scenario and values of jitter while adopting the former specification for the DCC facility layer [1] yielded quite different experimental results. Using 6 seconds of jitter was enough to trigger a polarization between node states. This stems from the fact that the previous DCC specification stipulates that the vehicles 'initial state is Active with a rate of 2Hz. As a consequence of this relatively low rate, vehicles react by lowering their state to Relaxed .Figure 7 illustrates the case where node ID =164 and node ID =159 are initially in the active state when the simulation begins. When node 159 must read CL numerical value, it switches to the relaxed state to recover with 0.04s CAM generation period. Node 164 does the opposite rate adjustment, since the time gap between the two vehicles measurements was enough for this node to perceive that the minimal Channel Busy Ratio (CBR) exceeded 40%. The same pattern continues as other neighboring nodes react to the latest restriction by switching their states from active to relaxed, and so on. The probability and the speed with which this situation propagates greatly depends on the probability of gaining channel access by the vehicle that has just taken the decision to enter one of the extreme states (Relaxed or restrictive), and hence on the value of the minimal and maximal contention windows (CW).

The comparison between the two DCC versions highlighted the importance of reducing the set of time gaps between the nodes' CL measurements by reducing the lower limit of the CAM interval generation and thus the impact of changing the initial DCC state.

Shorter jitter values of were used in order to observe the effect of short time delays between CL measurements. As shown in Figure 4, the average channel load of all the vehicles in simulation (sharing the same Clear Channel assessment range

(CCA)), i.e. CL values during 100ms, present more oscillations with a jitter of 2s than 4s at the beginning of the simulations. The relatively low value of average channel load obtained with 4s of jitter can be explained by the progressively number of vehicles taking part in the simulation.

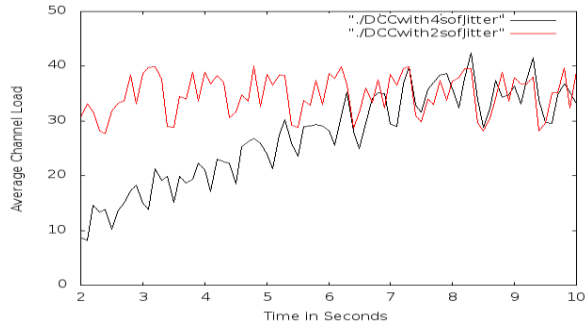

Fig.4 Average channel load oscillation with 2s and 4s of jitters at simulation start

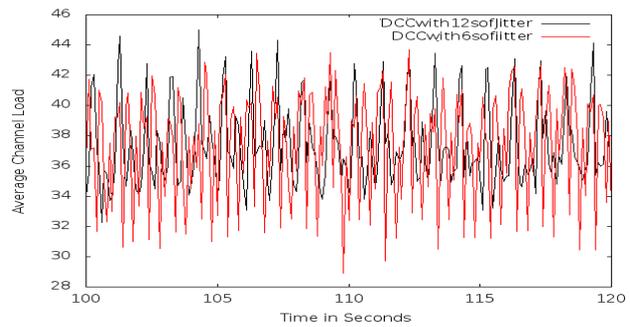

Fig.5 Average channel load oscillation with 6s and 12s of jitters at advanced simulation time

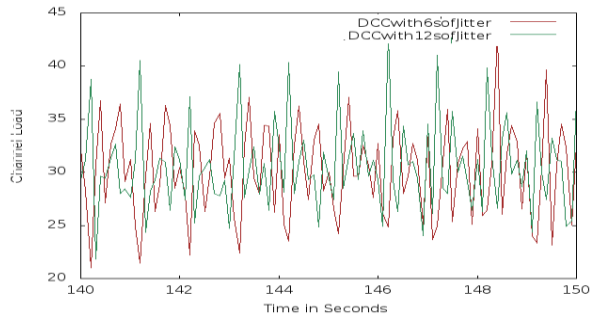

Fig.6 Channel load of a random node

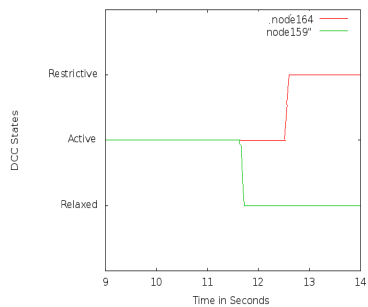

Fig.7 Polarization of states in former DCC version

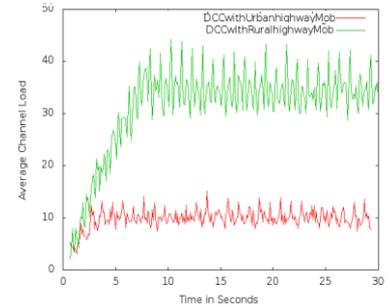

Fig.8 DCC Average Channel Load in Rural Highway and Urban Mobility Scenario

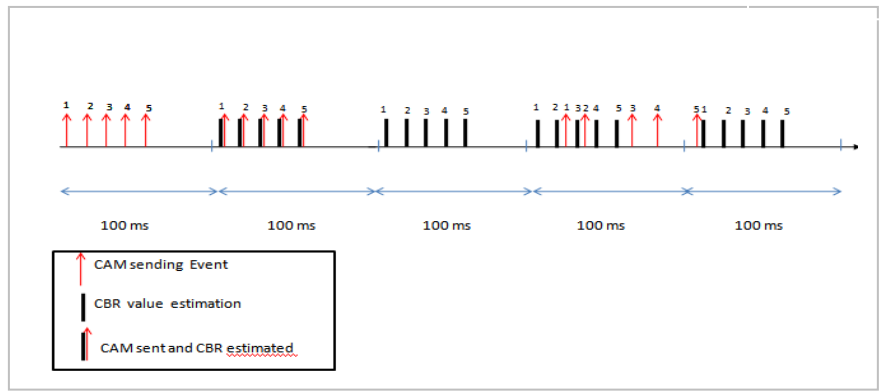

Fig.9 Illustration of CL oscillation reason in rural highway scenario

The average channel load starts rising considerably once the jitter value time has been reached. Even though the overall highest and lowest CBR values are quite similar during the first second of the simulations, the 4s jitter scenario presents a visible fluctuation in CL measurements as indicated in Figure.4. The oscillation effect of smaller jitter values is illustrated in later simulation times, as shown by Figure 5. Interestingly, this oscillation does not only take place in a global context, but also on a local scale, as is obvious on a randomly chosen vehicle ID (see Figure.6). In terms of state oscillation, the significant global and local CBR fluctuation when only the DCC latest facility layer specification is implemented does not really lead to DCC state oscillation. In fact, the sudden reduction and increase of CL under and above the two DCC machine thresholds does not imply a valid condition on the value of a minimal CL per 1s and maximal CL per 5s. To shed light on the notable CL fluctuation and we formulate two main hypotheses. The first is that oscillation may be due to the relatively low sampling rate for accessing the busy channel indication in Ns3 (1 Khz). The second assumption is that CAMs are generated according to dynamic conditions. Hence, in a rural and uniform highway scenario, vehicles tend to have a similar dominant dynamics patterns and time correlated reactions. For instance, when there is normal traffic on the road all the vehicles move smoothly and upon encountering a bottleneck they enter a deceleration step, and then resume their movement at a lower speed. A better illustration of CL oscillation in the context of an urban highway pattern is given in Figure 9. As we can see, the nodes read a CBR numerical value each 100ms. The number of nodes in the dense highway scenario is represented with 5 nodes. Each of the vehicles sends its first CAM message during the first 100 ms. After 100 ms has elapsed nodes conclude about CBR measurement supposedly equal to 40% for each one, as 4 busy time samples were detected out of a total of 10 samples. Assuming also that the CAM mobility condition was triggered for the nodes 5 CAM messages were sent by each one. Let us now suppose that front vehicles 1 and 2 suddenly started decelerating upon approaching a congested highway exit. This would lead to the same ratio of channel busy time for the next 100 ms but would not lead to a CAM sending event. The CBR value for the subsequent interval will be 0, 0, 10, 20 and 20 for each of the five nodes respectively, which leads to an average value of the CL that drops from 40 to 10. On the other hand, according to the hypothesis, vehicles sharing the CCA range in an urban multi lane and multi intersection scenario might present a higher variation in gaps between nodes stopping at traffic lights, maneuvering a turn, moving on a single congested lane, or on a fast lane, which will also differ local CBR perception and so present less risk of oscillation. In order to investigate our hypothesis, we conducted a simulation of a highway scenario with the same number of vehicles. We began by defining the turning and stopping probability for each group of nodes. Figure.8 shows that using the same amount of jitter time (4s) the average channel load for the urban scenario presents a lower value compared to the highway scenario. Oscillations are also reduced. Nevertheless, we still cannot confirm or reject any of the two hypotheses.

V. CONCLUSION AND FUTURE WORK

In this paper we discussed the network behavior under the latest ETSI DCC rate CAM control specification, in order to investigate its performance in terms of effectiveness, fairness, and stability, taking into account different jitter values, simulating a variety of penetration rates in dense networks. After observing the CL behavior we can confirm the influence on DCC stability of combining the CAM generation period with transmission power, queuing time and CCA threshold control. Thus, the stability aspects of the DCC algorithm should be further scrutinized taking into account the granularity of different mobility scenarios, in order to achieve robust improvements.